\documentclass{mystyle}
\usepackage{amsmath}
\usepackage{graphicx}
\usepackage{hyperref}
\usepackage{cite}
\usepackage{authblk}
\usepackage{textcomp}
\usepackage{indentfirst}
\usepackage{geometry}
\usepackage{caption}
\usepackage{float}

\title{Deep-learning-based electrode action potential mapping (DEAP Mapping) from annotation-free unipolar electrogram}
\author[1]{Hiroshi Seno}
\author[2,3]{Toshiya Kojima}
\author[1,4]{Masatoshi Yamazaki}
\author[1]{Ichiro Sakuma}
\author[2]{Katsuhito Fujiu}
\author[1,5]{Naoki Tomii}

\affil[1]{Graduate School of Engineering, The University of Tokyo, Tokyo, Japan}
\affil[2]{Department of Cardiovascular Medicine, The University of Tokyo, Tokyo, Japan}
\affil[3]{Department of Cardiovascular Medicine, Japanese Red Cross Medical Center, Tokyo, Japan}
\affil[4]{Department of Cardiology, Nagano Hospital, Soja, Japan}
\affil[5]{Research Center for Advanced Science and Technology, The University of Tokyo}

\begin{document}

\maketitle

\section*{Abstract}
\setlength{\parindent}{0em}

\textbf{Background and Aims}\\
Catheter ablation has limited therapeutic efficacy against non-paroxysmal atrial fibrillation (AF), and electrophysiological studies using mapping catheters have been applied to evaluate the AF substrate. However, many of these approaches rely on detecting excitation timing from electrograms (ECGs), potentially compromising their effectiveness in complex AF scenarios. Herein, we introduce Deep-learning-based Electrode Action Potential Mapping (DEAP Mapping), a deep learning model designed to reconstruct membrane potential images from annotation-free unipolar ECG signals.
\vspace{0.5\baselineskip}

\textbf{Methods}\\
We conducted ex vivo experiments using porcine hearts (N = 6) to evaluate the accuracy of DEAP Mapping by simultaneously performing fluorescence measurement of membrane potentials and measurements of epicardial unipolar ECGs. Membrane potentials estimated via DEAP Mapping were compared with those measured via optical mapping. We assessed the clinical applicability of DEAP Mapping by comparing the DEAP Mapping's estimations from clinically measured catheter electrode signals with those from established electrode-mapping techniques.
\vspace{0.5\baselineskip}

\textbf{Results}\\
DEAP Mapping accurately estimated conduction delays and blocks in ex vivo experiments. Phase variance analysis, an AF substrate evaluation method, revealed that the substrate identified from optical mapping closely resembled that identified from DEAP Mapping estimations (structural similarity index of \textgreater 0.8). In clinical evaluations, DEAP Mapping estimation observed several conduction delays and blocks that were not observed with existing methods, indicating that DEAP Mapping can estimate excitation patterns with higher spatiotemporal resolution.
\vspace{0.5\baselineskip}

\textbf{Conclusion}\\
DEAP Mapping has a potential to derive detailed changes in membrane potential from intra-operative catheter electrode signals, offering enhanced visualisation of the AF substrate from the estimated membrane potentials.

\begin{figure}[H]
  \centering
  \includegraphics[width=0.95\textwidth]{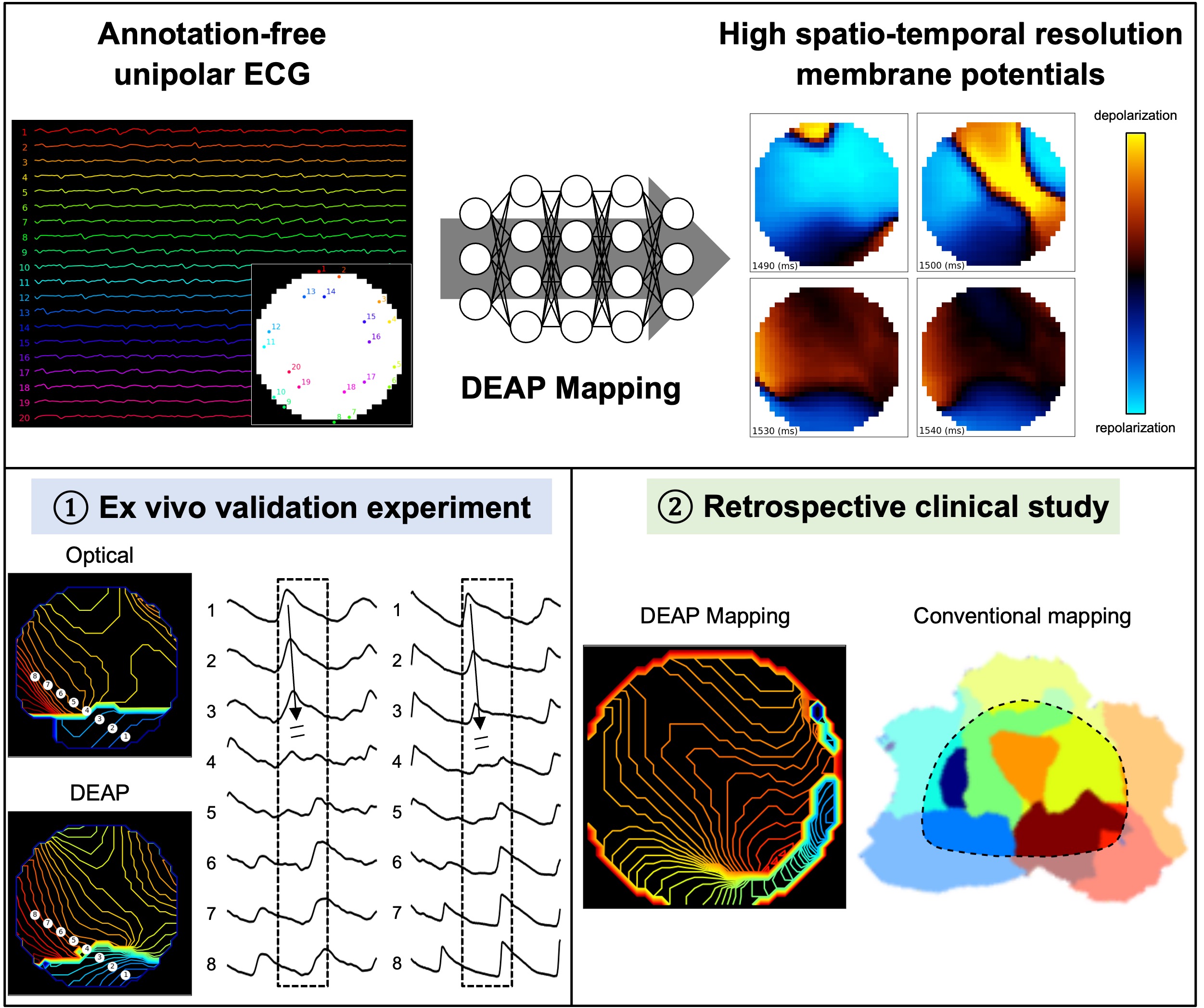}
  \label{fig:graphical_abstract}
\end{figure}

\section{Introduction}
\setlength{\parindent}{1em}

Atrial fibrillation (AF) is the most common tachyarrhythmia, affecting an estimated 1\%-2\% of the global population\cite{Andrade2014}. AF can elevate the risk of stroke and increase the rate of mortality in affected individuals\cite{Hart2001} and is believed to contribute to a decline in quality of life and cognitive impairments\cite{Thrall2006, Kalantarian2013}. Moreover, AF is predominantly found in older individuals, and considering the ageing population, the importance of AF treatment is consequently increasing\cite{Zoni2014, Lippi2021}.

Catheter ablation is widely employed for treating AF, and various ablation strategies have been examined to enhance its efficacy. A notable technique is pulmonary vein isolation (PVI), proposed by Haissaguerre et al.\cite{Haissaguerre1998}. PVI can be particularly effective against paroxysmal AF (PAF) and is now recognised as a standard ablation strategy\cite{Calkins2018}. However, PVI alone may not be sufficient for non-paroxysmal AF (non-PAF)\cite{Latchamsetty2014}, which is partly because the sources and substrates of AF can extend beyond the pulmonary veins. Over recent decades, many electrophysiological studies have been used to evaluate the substrate causing non-PAF. A pioneering approach was the CFAE mapping proposed by Nademanee et al., which emphasised the complexity of catheter electrode signals\cite{Nademanee2004}. However, recent clinical studies have not reported any significant therapeutic benefit for non-PAF with CFAE mapping\cite{Providencia2015}. Recently, the reconstruction of excitation patterns from electrode signals has been explored to determine ablation targets based on the interpretation of these excitations. An example is the FIRM mapping proposed by Narayan et al.\cite{Narayan2012}. In recent years, methods have been developed for estimating local excitation patterns using mapping catheters equipped with high-density electrodes, such as CARTO Finder by Biosense Webster Inc.\cite{Calvo2017} and ExTRa Mapping by Nihon Kohden Corp\cite{Sakata2018}. In these methods, the excitation timing is detected from the electrode signals, and by spatially interpolating the elapsed time from the excitation, the excitation pattern is visualised, from which the AF substrate is evaluated. However, during AF, complex abnormal excitations occur in both time and space, producing complex electrode waveforms that are distinct from the spike-like waveforms measured during sinus rhythm or tachycardia. These complex waveforms can be observed with both unipolar and bipolar electrodes. Consequently, accurately detecting the timing of excitation passage during AF is challenging, and the fundamental estimation of excitation patterns, which underlies the evaluation of fibrillation substrates, may be imprecise. Indeed, multi-centre clinical studies, verifying the efficacy of FIRM, have reported limited therapeutic effects for AF\cite{Buch2016}. Therefore, a reliable treatment for non-PAF has yet to be realised, emphasising the need for more effective methods of evaluating the AF substrate.

Various analytical methods for evaluating the AF substrate have been proposed in fundamental research areas where detailed membrane potential information can be obtained through fluorescent measurements. A prominent example is using isochronal maps where abnormal regions of excitation conduction are visualised as functional block lines\cite{Kodama2005, Amino2023}. However, the accuracy of substrate estimation with this method may be compromised, especially during complex AF excitations as this also relies on the detection of excitation timings. Consequently, recent research has proposed phase variance analysis for substrate evaluation during AF\cite{Tomii2015}. Phase variance analysis assesses the spatial dispersion against phase maps calculated from membrane potential movies and has potential to quantitatively detect abnormal excitation conduction regions demonstrated\cite{Tomii2021, Yamazaki2022}. Therefore, if detailed membrane potential information can be obtained from catheter electrode signals during ablation treatment, an AF substrate for non-PAF could be visualised using phase variance analysis.

Herein, we propose the Deep-learning-based Electrode Action Potential Mapping (DEAP Mapping), a deep learning model that reconstructs high spatiotemporal resolution membrane potential videos from annotation-free unipolar ECG. We conducted ex vivo experiments with excised porcine hearts to evaluate the membrane potential estimation accuracy using the proposed method and the fibrillation substrate estimation accuracy based on the estimated membrane potentials. Furthermore, to verify the clinical applicability of the proposed method, we applied this to catheter electrode signals measured during AF in patients with non-PAF and compared the estimated results with those of existing clinical mapping methods.

\section{Materials and Methods}
We developed DEAP Mapping, which predicts the temporal changes in membrane potentials within the measured area from annotation-free unipolar ECG signals (Figure 1). To verify the accuracy of the membrane potential distribution estimation via DEAP Mapping and the AF substrate evaluation based on the estimated membrane potentials, we established an experimental system that synchronises the measurement of extracellular potentials by electrodes with the fluorescent measurement of membrane potentials with a high-speed camera on excised porcine hearts (Figure 2A). Measured membrane potentials were compared with the estimated membrane potentials inferred from electrode signals using DEAP Mapping; phase variance indices calculated from the measured and estimated membrane potentials were then compared.

Ex vivo experiments using excised porcine hearts were approved by the Ethics Committee of The University of Tokyo (KA21-9), and our experiments were conducted in compliance with Directive 2010/63/EU of the European Parliament on the protection of animals used for scientific purposes and the NIH Guide for the Care and Use of Laboratory Animals. Pigs (N = 6) were anesthetized with ketamine hydrochloride, xylazine, atropine, propofol and isoflurane. To verify the adequacy of anaesthesia and analgesia, periodic assessments were conducted, including monitoring body movements and changes in vital signs in response to mild pain stimuli. After thoracotomy, hearts were rapidly excised. In this study, euthanasia was performed by exsanguination under deep anaesthesia followed by cardiac excision. The excised hearts were perfused via the aorta with modified Krebs–Ringer solution (pH 7.4, 95\% O\(_2\), 5\% CO\(_2\) and 36\textdegree C–38\textdegree C) at a flow rate of 200 mL/min using a Langendorff perfusion apparatus. Electrode catheters (Livewire, St. Jude Medical Inc., USA) were placed in the left atrium and coronary sinus, and stimulation and measurement were performed

\begin{figure}[H]
  \centering
  \includegraphics[height=0.85\textheight]{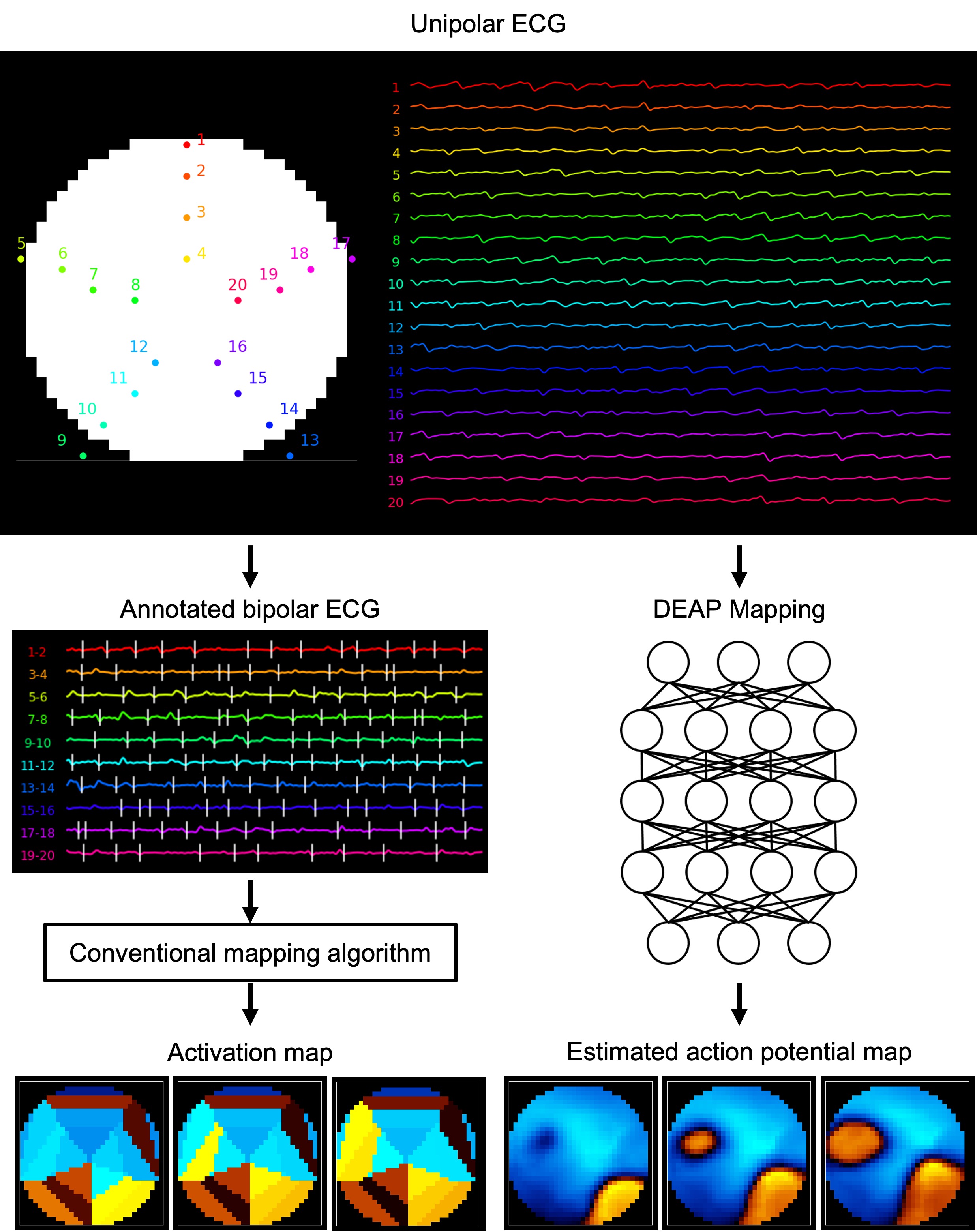}
  \caption{
    \textbf{Comparison between existing electrode mapping and the proposed DEAP Mapping method} Most prior studies related to electrode mapping detected excitation passage timing through peak detection of bipolar electrode signals between neighbouring electrodes. This approach involves spatial interpolation of elapsed times from the excitation passage to estimate excitation patterns. However, accurately detecting the timing of excitation passage during the fibrillatory complex excitations can be challenging, potentially reducing the accuracy in estimating the excitation patterns. In the current study, a deep learning model (Deep Electrode Action Potential Map: DEAP Mapping) was proposed that estimates changes in membrane potential within the measurement area from annotation-free unipolar signals of catheter electrodes without performing annotation processing to detect the timing of excitation passage.
  }
  \label{fig1}
\end{figure}
\newpage

\setlength{\parindent}{0em}
using a stimulator (SEC-5104, Nihon Kohden, Corp., Japan) and a polygraph (RMC-5000, Nihon Kohden Corp, Japan). An acute pressure-loaded AF model was created by applying water column pressure inside the atrium as described in our previous studies \cite{Yamazaki2012, Hori2023}, with a 12-cm pressure applied in this study. AF was induced by high-frequency pacing using an electrode catheter placed inside the left atrium.

\setlength{\parindent}{1em}
Electrode measurements were conducted in parallel with fluorescent measurements of membrane potentials, and the electrode measurements were performed without obstructing the optical measurements. To enable this, we created transparent electrode arrays embedded with Ag–AgCl electrodes in polydimethylsiloxane (PDMS). The electrode array designs were modelled after the clinical catheter electrode configurations, with a pentagon design comprising four electrodes on each of the five spines and a spiral design with 20 electrodes arranged in a spiral pattern (Figure 2B). The transparent PDMS electrodes were placed in contact with the epicardial side of the left atrial appendage (Figure 2C), and the unipolar potentials with reference to the aorta were acquired using the polygraph.

Fluorescent measurements of membrane potentials were performed using an optical mapping technique, which uses a voltage-sensitive dye and a high-speed camera, similar to that used in a previous study \cite{Yamazaki2007}. Heart models were perfused with modified Krebs–Ringer solution containing 0.1 \(\mu\)M of the voltage-sensitive dye Di-4-ANEPPS and 19 mM 2,3-butanedione monoxime (BDM) to suppress muscle contraction-caused motion artefacts. AF was induced by high-frequency pacing while the custom electrode array was in contact with the left atrial appendage. Subsequently, excitation light from an LED-ring light at 520 nm (Xeon 3 Power Pure Green LED, OptoSupply, China) was shined on the target left atrial appendage, and the fluorescence passing through a high-pass optical filter (\textgreater 600 nm, R-60, Nikon, Japan) was captured by a high-speed camera (FASTCAM Mini-AX50, Photron, Japan) at 1000 fps, 512 \(\times\) 512 pixels. Synchronisation of the optical and electrical measurements was achieved by initiating the high-speed camera recording with an external trigger device and simultaneously inputting the trigger signal into the polygraph.

The measured electrode waveforms were used as input into DEAP Mapping to estimate the changes in membrane potential occurring within the measurement area. For comparison, activation maps were created by detecting the excitation passage timing from the electrode waveforms through peak detection and spatially interpolating the elapsed time from these timings. Using affine transformations based on the electrode positions captured via camera measurements, the corresponding optical measurements were extracted for the mapping area. These extracted optical results served as the basis for evaluating the mapping performance of the proposed DEAP Mapping and activation maps.

To assess the clinical applicability of the DEAP Mapping, we conducted a retrospective clinical study using catheter electrode signals obtained from patients with non-PAF. The retrospective study was approved by the Research Ethics Committee of the Faculty of Medicine of the University of Tokyo (2023232NI) and was conducted in accordance with the principles of the Declaration of Helsinki. For this study, intracardiac ECGs recorded in the left atrium during AF episodes from non-PAF patients (N = 2) treated at the University of Tokyo Hospital were utilized. In this study, informed consent was obtained through an opt-out approach. Participants were informed about the study via public notices and were given the opportunity to decline participation. The CARTO3 system (Biosense Webster Inc., USA) served as the three-dimensional navigation system in these instances. Either a Pentaray catheter (Biosense Webster Inc., USA) for CARTO Finder or Reflexion catheter (St. Jude Medical Inc., USA) for ExTRa Mapping was employed as the mapping catheter. The position of the mapping catheters was located using the CARTO3 system, and ECG recordings were performed with a polygraph (RMC-5000, Nihon Kohden Corp, Japan). Retrospectively, excitation patterns were estimated from the measured ECGs via DEAP Mapping, and the estimation was compared with the predictions from CARTO Finder and ExTRa Mapping.

\section{Results}
A total of 55 optical and electrode simultaneous measurements were performed on six excised porcine heart models. Of these, 10 cases were excluded because of optical measurement abnormalities caused by ischaemia or tissue damage or because of photobleaching of voltage-sensitive dye, four cases were excluded because of

\begin{figure}[H]
  \centering
  \includegraphics[height=0.8\textheight]{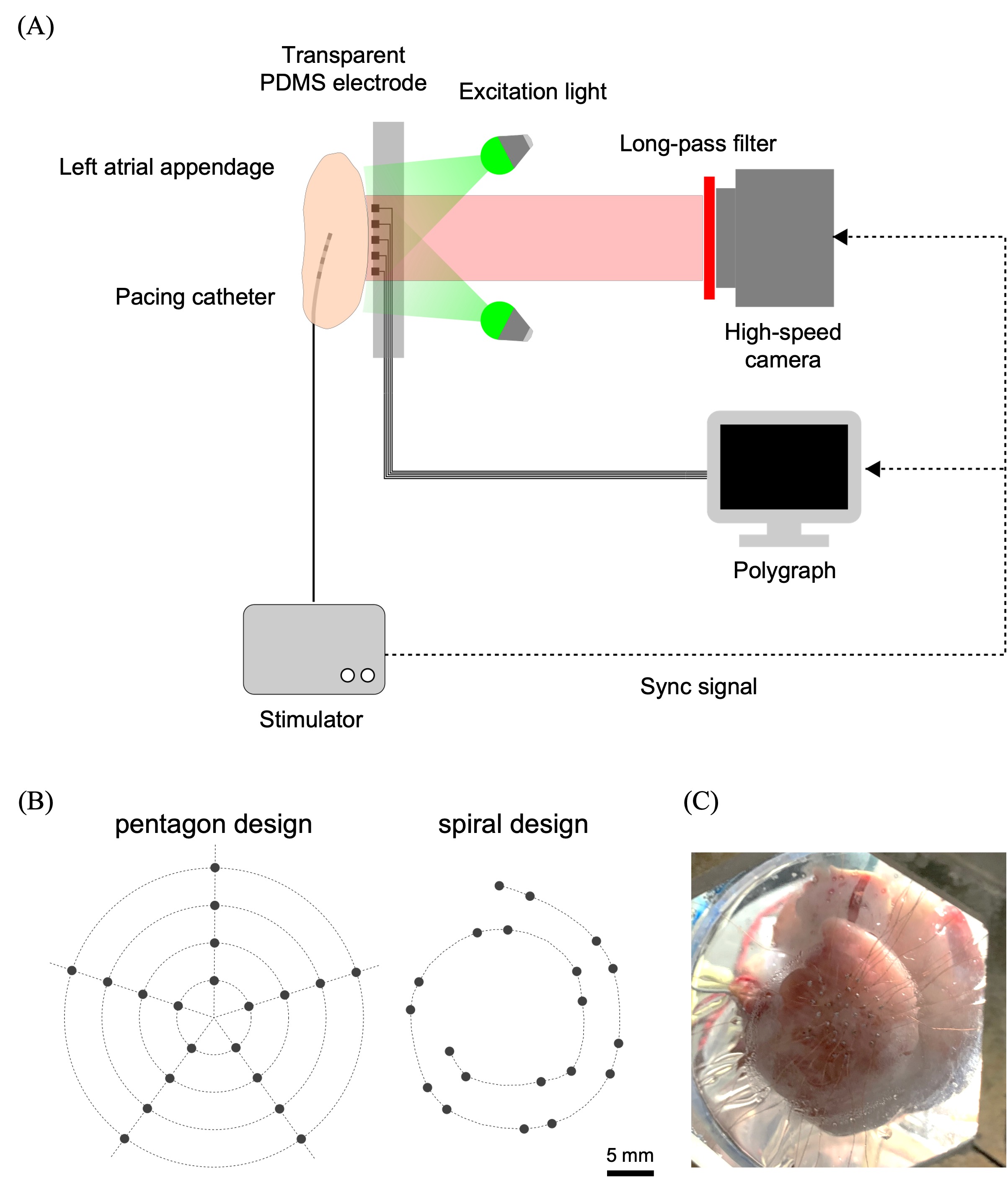}
  \caption{
    \textbf{Overview of the ex vivo experiment} (A) Conceptual diagram of the simultaneous optical and electrode measurement system. Excised porcine hearts (N = 6) were used. AF was induced in an acute pressure-loaded model via high-frequency pacing from an electrode catheter placed inside the left atrium, and the measurement target was set on the epicardial side of the left atrial appendage. Electrode signals were measured using a polygraph. The heart model was stained with a voltage-sensitive dye, and optical measurement of the membrane potential was performed using a high-speed camera. The synchronisation of electrode and optical measurements was achieved by externally initiating the high-speed camera recording and simultaneously inputting the trigger signal into the polygraph. (B) Electrode patterns used in the ex vivo experiment. Electrode configurations were designed based on those used in clinical catheter electrodes, creating two types: a pentagon design with four electrodes on each of five spines, and a spiral design with 20 electrodes arranged in a spiral pattern. (C) Experimental setup. Custom-made transparent resin electrodes were placed in contact with the epicardial side of the left atrial appendage for electrode measurement. Simultaneously, fluorescence passing through the transparent electrodes was measured, allowing for membrane potential measurement.
  }
  \label{fig2}
\end{figure}
\newpage

\setlength{\parindent}{0em}
electrode measurement abnormalities caused by poor contact and seven cases were excluded because of atrial tachycardia with excitation cycles of \textgreater 200 ms. The remaining 34 measurements were included in the analysis. In this paper, we present several representative analysis results, with other additional analytical examples provided in the supplemental information.

\setlength{\parindent}{1em}
Figures 3 and 4 illustrate a comparison between the inferred membrane potentials by the proposed DEAP Mapping and activation maps against the membrane potential videos obtained from optical mapping. In these figures, Panel A represents the unipolar electrode waveforms around the time of the excitation pattern comparison; the time period enclosed by the white dashed line corresponds to the time frame during which the membrane potential distribution is compared in Panels B–D. Panel B shows optical mapping results within the electrode-mapping area, Panel C shows the time series changes in the membrane potential inferred by DEAP Mapping and Panel D shows the activation maps obtained by spatial interpolation of the excitation passage timings detected from the electrode signals.

Figure 3 presents an example using the pentagon electrode configuration. The optical measurement results (Figure 3B) showed that an excitation entered from the lower right region (765 ms) and a clockwise spiral excitation occurred in the upper right area in this case (780–830 ms). Upon examining the inferred membrane potentials by DEAP Mapping (Figure 3C), excitation occurring in the lower right region and circulating toward the upper right area was estimated with high spatiotemporal resolution by DEAP Mapping. However, from Figure 3D, the activation map based on excitation passage timing provided a rough estimate of the excitation patterns similar to Figures 3B and C, but with lower spatial resolution when compared with the optical measurement and DEAP Mapping.

Figure 4 shows an example of simultaneous measurements using the spiral electrode configuration. As seen in Figure 4B, after the excitation propagation originating in the lower area was blocked (1460 and 1490 ms), a new excitation was observed to propagate from the upper right to the lower left (1500–1540 ms). As shown in Figures 4C and D, similar to Figure 3, the membrane potential changes inferred by DEAP Mapping resembled the excitation patterns observed in optical measurements and exhibited superior spatiotemporal resolution compared with the activation map.

Figure 5 assesses the accuracy of the AF substrate estimation. Figures 5A and B present an example using the pentagon design, and Figures 5C and D present an example using the spiral design. These figures show a comparison of the changes in membrane potential by visualising the isochronal map and phase variance indices for the cases shown in Figures 3 and 4. In Figures 5A and C, the isochronal maps based on optical measurement results revealed a clockwise rotating excitation in the upper right area of Figure 5A and an excitation conduction block in the lower area of Figure 5C. The isochronal maps drawn from the membrane potentials inferred by DEAP Mapping represent the rotating excitation and conduction block observed in the optical measurement isochronal maps, indicating a similarity. Conversely, the isochronal maps drawn based on the activation map were strongly influenced by the boundaries of spatial interpolation, producing lower spatial resolution and making it difficult to grasp the excitation patterns compared with the optical measurements and DEAP Mapping isochronal maps. In the accuracy evaluation of the estimated sites of abnormal excitations using phase variance analysis, the phase variance indices calculated from the membrane potentials obtained by DEAP Mapping had a high image similarity to the phase variance indices obtained from optical measurements for both electrode configurations. However, the phase variance indices drawn from the activation map showed high phase variance regions at the boundaries of spatial interpolation, which significantly differed from those obtained by optical measurements and DEAP Mapping. Figures 5B and 5D represent the membrane potential changes at eight points displayed on the isochronal maps of Figures 5A and 5C, respectively, with the dotted line–enclosed area corresponding to the timeframe of the isochronal maps. Thus, DEAP Mapping could estimate the rotating excitation wave centred in the upper right area in the example shown in Figures 5A and 5B and the excitation conduction block in the lower area in Figures 5C and 5D, similar to optical measurements, indicating accurate estimation at the membrane potential level. Figure 5E shows the results of assessing the structural similarity (SSIM) of phase variance indices calculated from the DEAP Mapping and activation map with reference to phase variance indices obtained from optical measurements (pentagon design: 22 cases, spiral design: 12 cases). In Figure 5E, the left panel shows a scatter plot with the horizontal axis representing the SSIM between the phase variance index obtained from the activation map and the optical measurement, and the vertical axis representing the SSIM between the

\begin{figure}[H]
  \centering
  \includegraphics[height=0.8\textheight]{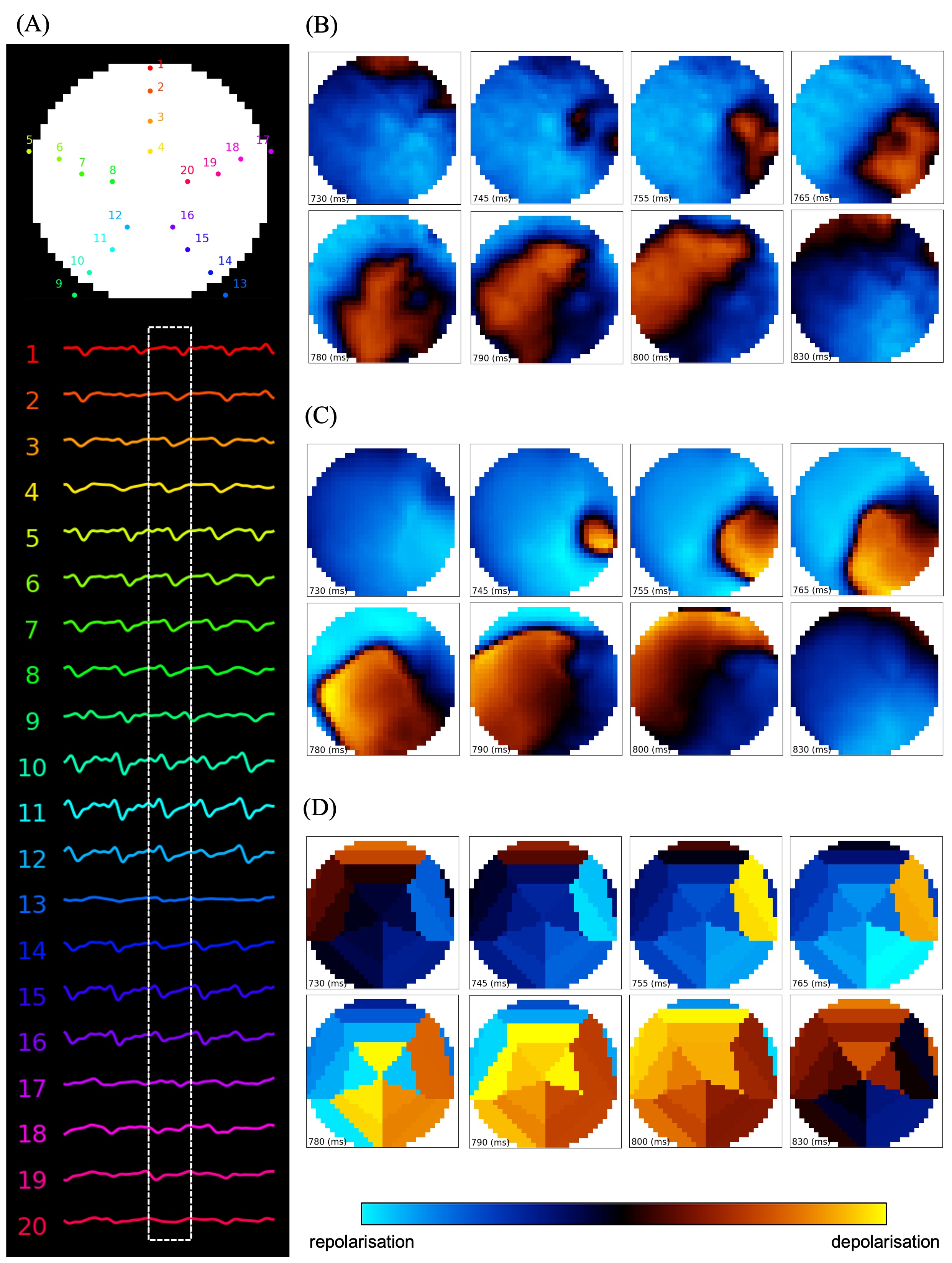}
  \caption{
    \textbf{Membrane potential estimation results in ex vivo experiment (pentagon electrodes)} (A) Unipolar electrode signals from each electrode around the time of excitation pattern comparison. The period enclosed by the white dotted line corresponds to the timeframe during which the membrane potential distribution was compared in Panels B-D. (B) Membrane potential obtained from optical measurement. During the observed timeframe, excitation entered from the lower right of the area (765 ms), and a clockwise rotating excitation occurred in the upper right region (780-830 ms). (C) Membrane potential inferred by DEAP Mapping from the unipolar electrode signals. DEAP Mapping could estimate with high spatiotemporal resolution the emergence of excitation in the lower right area and the rotating excitation in the upper right, as observed in the optical measurement (Figure B). (D) Activation map obtained by detecting the excitation passage timing from the rise in electrode signals and spatially interpolating the elapsed time from the excitation passage. A similar pattern of excitation could be observed as in Panels B and C, where the spatial resolution was lower compared with that of the optical mapping and DEAP Mapping.
  }
  \label{fig3}
\end{figure}
\newpage

\begin{figure}[H]
  \centering
  \includegraphics[height=0.8\textheight]{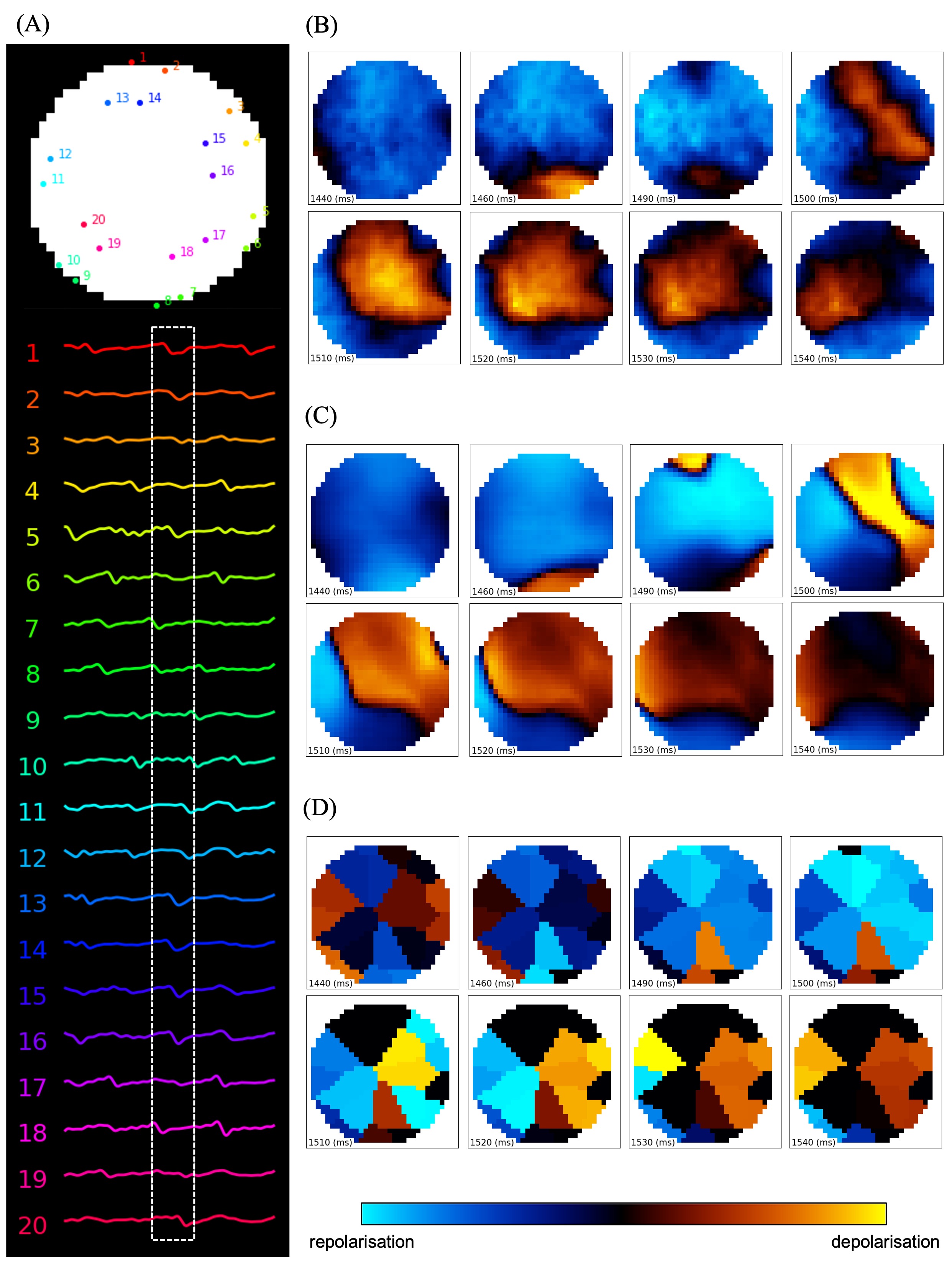}
  \caption{
    \textbf{Membrane potential estimation results in ex vivo experiment (spiral electrodes)} (A) Unipolar electrode signals from each electrode around the time of excitation pattern comparison. The period enclosed by the white dotted line corresponding to the timeframe during which the membrane potential distribution was compared in Panels B-D. (B) Membrane potential obtained from optical measurement. Following blocking of excitation propagation originating in the lower area (1460, 1490 ms), a new excitation propagated from the upper right to the lower left area (1500-1540 ms). (C) Membrane potential inferred by DEAP Mapping from the unipolar electrode signals. Similar to the example in Figure 3, DEAP Mapping could estimate the excitation pattern similar to that observed in optical measurements. (D) Activation map obtained by detecting the excitation passage timing from the rise in electrode signals and spatially interpolating the elapsed time from the excitation passage. As in Figure 3, while a similar pattern of excitation was observed in Panels B and C, the spatial resolution was lower compared with that of the optical mapping and DEAP Mapping.
  }
  \label{fig4}
\end{figure}
\newpage

\setlength{\parindent}{0em}
phase variance index obtained from the DEAP Mapping and the optical measurement. This scatterplot shows that the SSIM for phase variance index estimation based on the activation map was broadly distributed from approximately 0.2 to 0.8, whereas the SSIM for all estimated measurements based on DEAP Mapping was \textgreater 0.6. Furthermore, across all cases, the SSIM for DEAP Mapping was higher than that for the activation map. The violin plots for the SSIM based on the activation map and DEAP Mapping for pentagon and spiral electrode design are shown in Figure 5F. Furthermore, irrespective of the electrode design, the average SSIM for the phase variance index based on the activation map was \textless 0.4, whereas when using DEAP Mapping, the average SSIM was \textgreater 0.8, indicating that DEAP Mapping improves the estimation of the phase variance index.

\setlength{\parindent}{1em}
Figures 6 and 7 shows the comparison results between DEAP Mapping and the existing electrode-mapping techniques in retrospective clinical study. Each figure in Panel A depicts the atrial shape and measurement locations acquired on the CARTO system. Panel B (left) illustrates the relationship between the region of interest in the DEAP Mapping and electrode positions, and the right image in Panel B represents the evaluation results of the phase variance index for this measurement case. Panel C shows the unipolar electrode signals measured with the polygraph, comparing the excitation pattern estimation results for the period enclosed by the dotted line. Panel D shows the inferred membrane potential results obtained by DEAP Mapping and the isochronal map while Panel E represents the excitation pattern estimation results obtained by the existing electrode mapping. The dotted line–enclosed areas correspond to the visualisation region of DEAP Mapping.

CARTO Finder and DEAP Mapping are compared in Figure 6, where the measurement area was the upper posterior wall of the left atrium of patient 1. We observed that during the time of interest, a clockwise rotating excitation occurred within the measurement region, and a conduction block was inferred via DEAP Mapping to have occurred in the lower right area (Figure 6D). The excitation conduction estimation results by CARTO Finder and the activation map calculated based on the excitation passage timing are shown in Figure 6E. A clockwise spiral excitation similar to the inference by DEAP Mapping was observed in the region of interest enclosed by the dotted line. However, the conduction block in the lower right area observed in the DEAP Mapping inference was not observed, inferring that circumferential spiral excitation had occurred instead.

ExTRa Mapping and DEAP Mapping are compared in Figure 7. This example is the result of measuring the lower posterior wall of the left atrium of patient 2. DEAP Mapping inferred the propagation of a planar excitation wave from the lower to the upper area, with excitation conduction delay and wave breaks occurring in the upper region during the propagation process (Figure 7D). For the excitation estimation results obtained via ExTRa Mapping (Figure 7E), the isochronal map (on the right side of the figure) was obtained by superimposing the wavefront represented by the white line within ExTRa Mapping. ExTRa Mapping predicted a planar excitation wave passing from the lower to the upper area, similar to DEAP Mapping. However, the conduction block and wave break in the upper area observed in DEAP Mapping were not clearly visualised.

\section{Discussion}
The result from ex vivo experiments conducted with simultaneous electrical and optical measurements revealed that the proposed DEAP Mapping method offers significantly higher accuracy and spatiotemporal resolution in estimating excitation patterns compared to the conventional activation map, which is based on annotation of excitation passage timing (Figures 3, 4). The limitation in spatial resolution of the activation map is thought to be attributed to the spatial interpolation process of excitation passage timing. Generally, the accuracy of interpolation processes is strongly influenced by the number of sampling points, making it inherently challenging to map excitation patterns with high spatial resolution from sparsely placed and limited electrode potential signals. On the other hand, the reconstruction of high-dimensional information from low-dimensional data is known as super-resolution task in the field of deep learning. This research also leverages the advanced pattern recognition capabilities of deep learning models to achieve high-precision super-resolution, which was challenging using traditional spatial interpolation algorithms.

\begin{figure}[H]
  \centering
  \includegraphics[height=0.75\textheight]{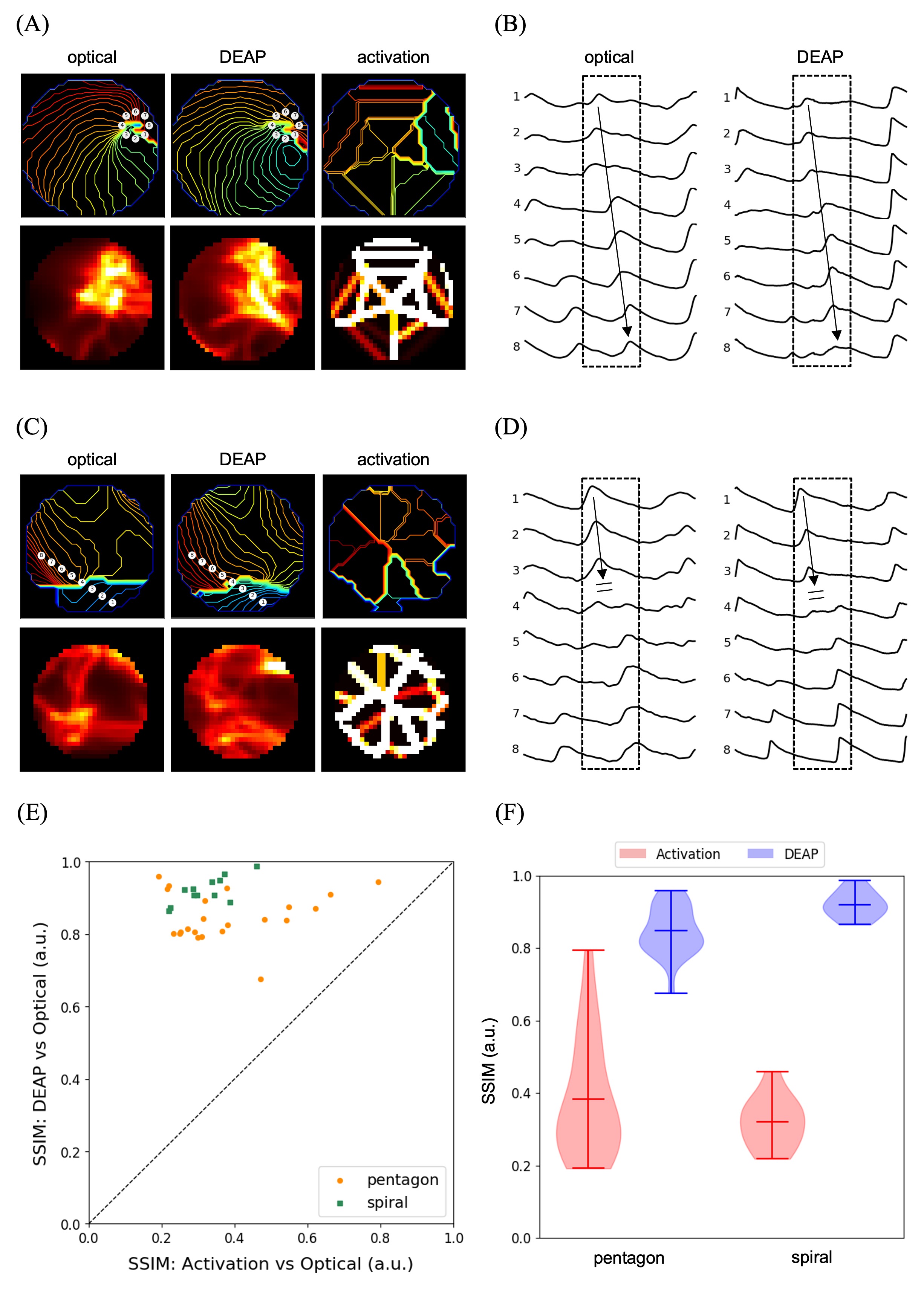}
  \caption{
    \textbf{Substrate estimation results in ex vivo experiment} (A) Comparison of isochronal maps and phase variance indices using pentagon electrodes. Isochronal maps and phase variance indices were compared for the same measurement example and time as in Figure 3. In both isochronal and phase variance maps, DEAP Mapping visualized excitation patterns similar to those measured by optical mapping, outperforming the activation map.  (B) Membrane potential changes at eight points displayed on the isochronal maps of optical measurement and DEAP Mapping from Figure (A). DEAP Mapping could estimate rotating excitation waves centred in the upper right area at the membrane potential level, similar to that estimated with optical measurements. (C) Comparison of isochronal maps and phase variance indices using spiral electrodes. Isochronal maps and phase variance indices were compared for the same measurement example and time as in Figure 4. For the spiral electrode configuration, both the isochronal maps and phase variance indices created from DEAP Mapping showed high similarity to those from optical measurements. (D) Membrane potential changes at eight points displayed on the isochronal maps of optical measurement and DEAP Mapping from Figure (C). DEAP Mapping could estimate the excitation conduction block in the lower area at the membrane potential level. (E) Evaluation of the accuracy of phase variance index estimation. The SSIM for DEAP Mapping was higher than that for the activation map across all cases. Furthermore, irrespective of the electrode design, the average SSIM for activation map was <0.4, whereas the average SSIM for DEAP Mapping was >0.8, indicating that the similarity was significantly improved.
  }
  \label{fig5}
\end{figure}
\newpage

\begin{figure}[H]
  \centering
  \includegraphics[height=0.75\textheight]{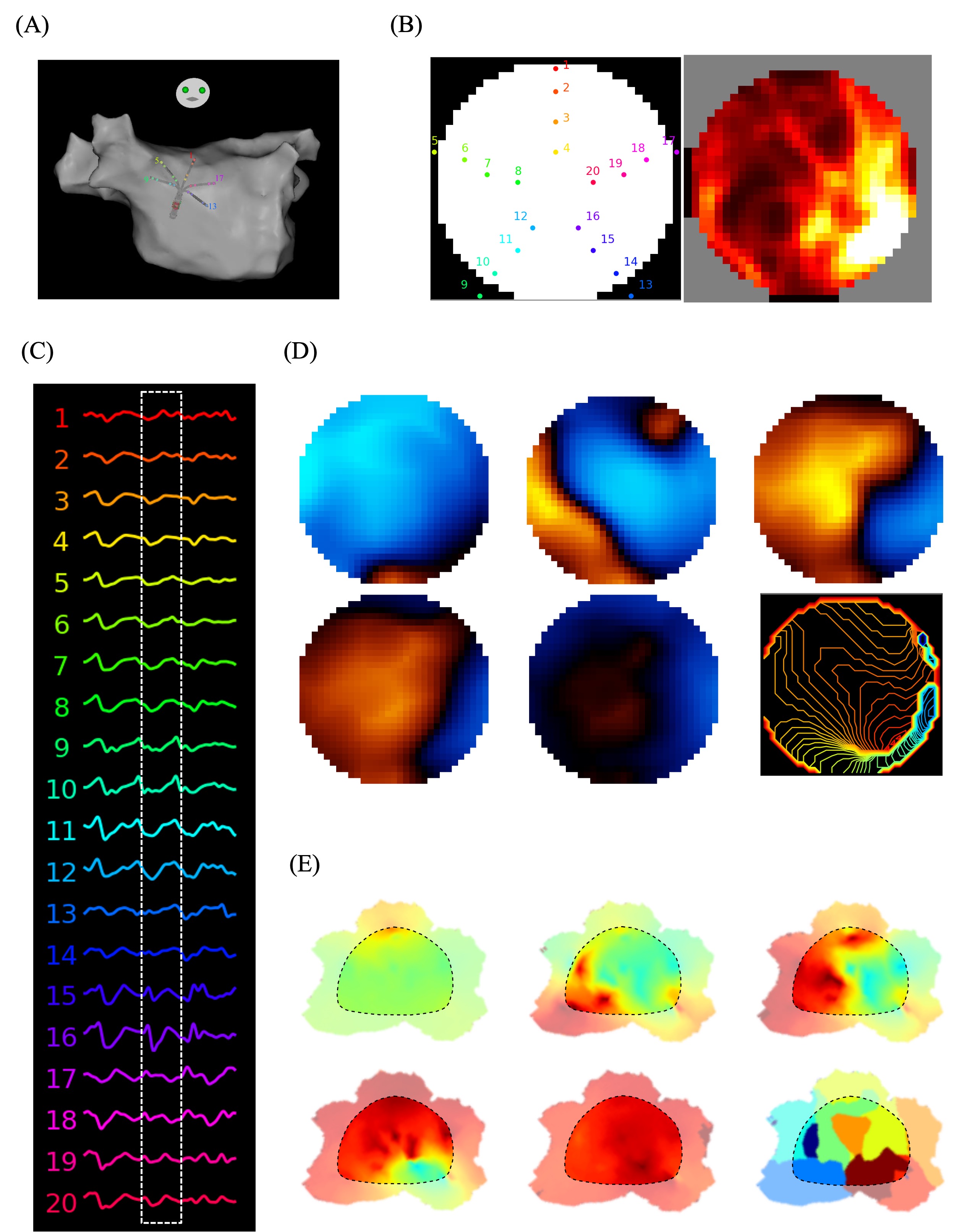}
  \caption{
    \textbf{Comparison with existing mapping method (CARTO finder) using clinical data} (A) Left atrial shape model and measurement location. Data measured with a pentaray catheter on the anterior wall of the left atrium were used for comparison between CARTO Finder and DEAP Mapping. (B) Electrode positioning relative to the region of interest (ROI) for DEAP Mapping and the estimated phase variance index. (C) Unipolar measurement signals from the pentaray catheter inputted into DEAP Mapping. The timeframe around the comparison of excitation patterns is shown, with the area enclosed by the white dotted line corresponding to the timeframe for comparing excitation estimation results. (D) Excitation estimation results via DEAP Mapping. This panel shows the time series of membrane potential changes inferred by DEAP Mapping and the isochronal map (bottom right). Clockwise spiral excitation occurred within the measurement area, and a conduction block occurred in the lower right area, causing the disappearance of the overall excitation as inferred by DEAP Mapping. (E) Excitation estimation results by CARTO Finder showing the excitation conduction estimation by CARTO Finder and the activation map (bottom right) calculated based on the timing of excitation passage; the dotted line–enclosed area corresponds to the ROI of DEAP Mapping. Thus, the clockwise spiral excitation could be estimated, similar to the inference by DEAP Mapping. However, the conduction block in the lower right area observed in the DEAP Mapping inference was not seen, and CARTO Finder inferred that a circumferential spiral excitation had occurred.
  }
  \label{fig6}
\end{figure}
\newpage

\begin{figure}[H]
  \centering
  \includegraphics[height=0.75\textheight]{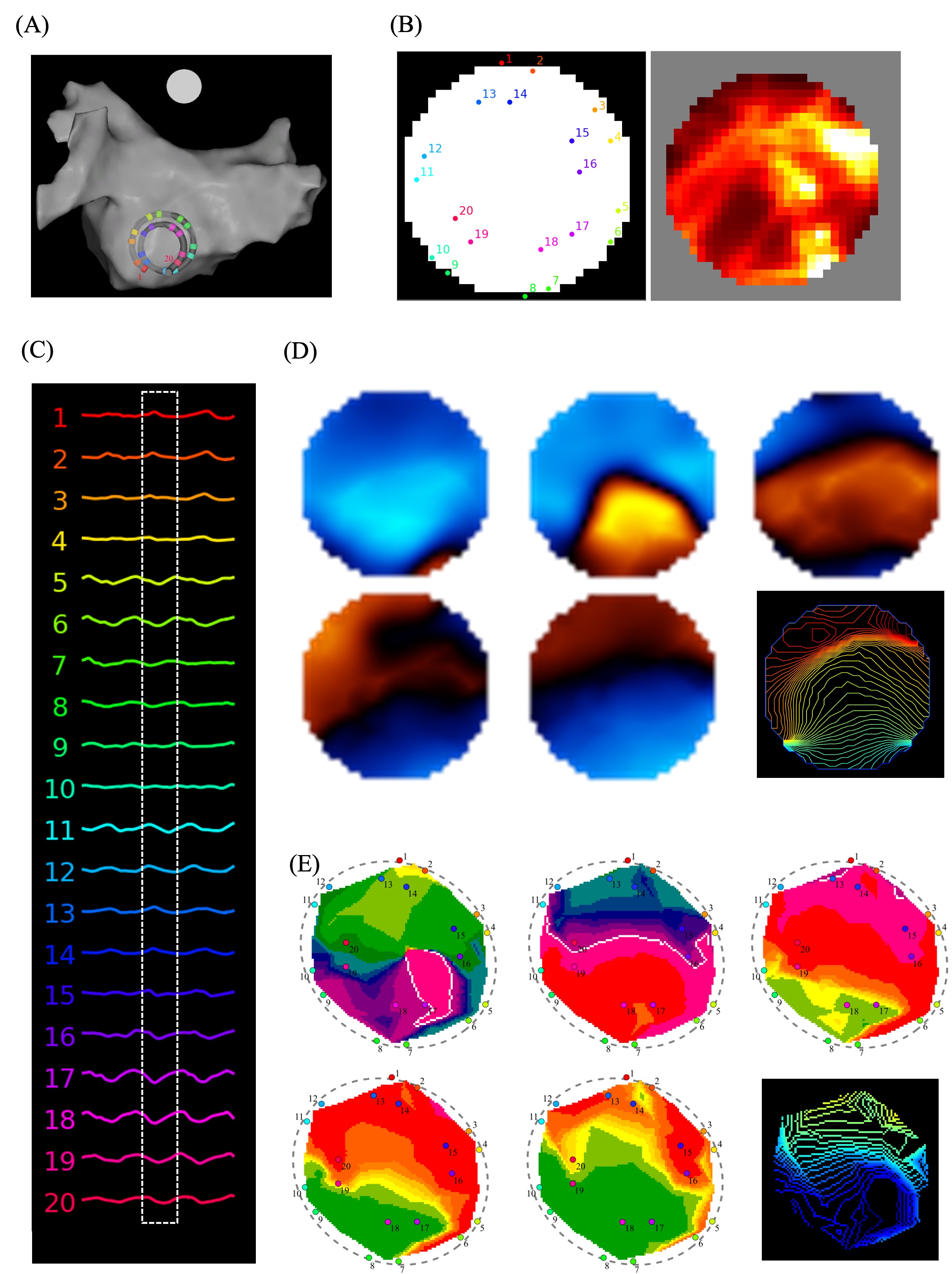}
  \caption{
    \textbf{Comparison with existing mapping method (ExTRa Mapping) using clinical data} (A) Left atrial shape model and measurement location. Data measured with a Reflexion catheter on the lower posterior wall of the left atrium were used for comparison between ExTRa Mapping and DEAP Mapping. (B) Electrode positioning relative to the ROI of the DEAP Mapping and the estimated phase variance index. (C) Unipolar measurement signals from the Reflexion catheter inputted into DEAP Mapping. The area enclosed by the white dotted line corresponds to the timeframe for comparing excitation estimation results. (D) Excitation estimation results via DEAP Mapping showing the time series of membrane potential changes inferred by DEAP Mapping and the isochronal map (bottom right). DEAP Mapping estimation of planar excitation wave propagation from bottom to top in the area, with excitation conduction delays and wave breaks occurring in the upper region during the propagation process. (E) Excitation estimation results via ExTRa Mapping showing the excitation conduction estimation via ExTRa Mapping and the isochronal map (bottom right) obtained by superimposing the wave front displayed on ExTRa Mapping; the dotted line–enclosed area corresponds to the ROI of the DEAP Mapping. ExTRa Mapping also predicted a planar excitation wave passing from bottom to top, similar to DEAP Mapping. However, the conduction block and wave breaks in the upper area observed in DEAP Mapping were not clearly visualised in ExTRa Mapping.
  }
  \label{fig7}
\end{figure}
\newpage

Furthermore, a significant difference was observed in estimation of arrhythmogenic substrate through phase variance analysis when using membrane potentials inferred by DEAP Mapping versus those obtained using activation map (Figure 5). The difference is thought to be because the accuracy of phase variance analysis is affected by the spatiotemporal resolution of the excitation conduction pattern estimation. Phase variance analysis focuses on the local variability of phases at the sites of abnormal excitation leading to the occurrence and maintenance of AF, such as conduction blocks and spiral excitation, analyzing arrhythmogenic substrates in complex fibrillatory excitation\cite{Tomii2015}. Prior studies on phase variance have employed membrane potential signals measured with sufficiently high spatiotemporal resolution, such as those from numerical simulations or optical mapping, and the impact of spatiotemporal resolution has not been fully evaluated\cite{Tomii2021, Yamazaki2022}. However, this study clearly shows that applying phase variance analysis to activation map, constrained by the number of electrodes, results in apparent variability of phases in the inter-electrode boundary regions, leading to artifacts in the phase variance indices as seen in Figures 5A and C. These findings suggests that accurate estimation of abnormal excitation areas through phase variance analysis requires excitation estimation with high spatiotemporal resolution, which is met by the DEAP Mapping's resolution.

Additionally, studies using catheter electrode signals measured from non-PAF patients indicated that membrane potential changes inferred by DEAP Mapping possess higher spatiotemporal resolution compared to existing mapping methods (Figures 6, 7). Unlike the ex vivo experiments, optical mapping could not be performed in these studies, and the accuracy of excitation pattern estimation could not be evaluated. However, as discussed above, the evaluation of abnormal excitation during AF through phase variance analysis may require excitation pattern estimation with high spatiotemporal resolution, and combining DEAP Mapping with phase variance analysis may allow for the evaluation of detailed abnormal excitation regions from catheter electrode signals in clinical practice.

In this study, we have several limitations. In ex vivo experiments, the measurement target was set on the epicardial side of the left atrium to realise synchronised optical and electrode measurements. However, in clinical settings, the endocardial side with a complex 3D structure is the measurement target, and the contact of the catheter electrodes with the heart wall may be insufficient, potentially altering the waveform characteristics. Although differences in measurement environment exist, it is believed that the clinical applicability of DEAP Mapping to actual clinical environment has been validated to some extent through retrospective clinical study utilizing potential signals recorded from non-PAF patients. Additionally, we used an acute pressure-loaded AF model by applying water column pressure to atria treated with an excitation–contraction uncoupler (BDM) to induce AF. Stretching stimulus from water column pressure and BDM administration can affect the electrophysiological behaviour of myocardial cells and heart tissue. Therefore, AF induced in the ex vivo model may differ in mechanisms of generation and persistence from that found in chronic non-PAF. However, as shown in Figures 3 and 4, the AF induced in our ex vivo model exhibited complex excitation patterns, including blockages and wave breaks, meeting the requirements of our study to validate mapping accuracy. Moreover, this study used clinical measurement data to render membrane potential changes and evaluate AF substrates using the phase variance index. Nonetheless, further research on ablation treatment strategies based on the visualised index is necessary to demonstrate the effectiveness of these indices for treating non-PAF.

\section{Conclusion}
Herein, with the goal of establishing effective treatment for non-PAF, we proposed a deep learning model that estimates the excitation patterns of a measured area from annotation-free electrode signals. We conducted ex vivo validation experiments using excised porcine heart models and preliminary investigations using clinical measurement data to assess the potential for membrane potential estimation and AF substrate estimation using the proposed method. Ex vivo experiments demonstrated the potential of the proposed method to predict membrane potentials with an accuracy comparable with that of optical mapping. Furthermore, preliminary experiments using clinical data showed that the proposed method can visualize excitation conduction delays and blocks, which were previously undetectable using existing electrode techniques. Thus, DEAP Mapping proposed in this study can potentially estimate detailed membrane potential changes in the measured area from intra-operative catheter electrode signals and visualise detailed AF substrates from the estimated membrane potentials.
\section*{Funding}
This work was supported by Japan Society for the Promotion of Science [21H04953 to I.S., 21K18036 to N.T., 23K07547 to M.Y.]; and Japan Agency for Medical Research and Development [JP23he0422020j0001 to N.T.].

\section*{Acknowledgements}
We extend our heartfelt gratitude to Dr. H. Tsukihara and Dr. A. Fujisawa for their invaluable cooperation and support in conducting the ex vivo experiments. Their expertise and guidance significantly contributed to the success of this study. We would also like to express our sincere appreciation to Mr. M. Tsuzuki for his unwavering support throughout this research endeavour. Furthermore, we are grateful to Mr. K. Yamamoto for his meticulous English proofreading of the manuscript, which greatly improved the clarity and quality of the final document.

\bibliographystyle{unsrt}
\bibliography{references}

\end{document}